\documentclass[epj]{svjour}
\usepackage{graphics}
\begin{document}
\title{The JLab Upgrade - Studies of the Nucleon with {\tt\bf CLAS12}.}
\subtitle{Selected topics }
\author{Volker D. Burkert}
%
%
\institute{Jefferson Lab, Newport News, Virginia, USA}
\date\today
%
\abstract{An overview is presented on the program to study the nucleon structure at the 12 GeV JLab upgrade using the {\tt CLAS12} detector. The focus is on deeply virtual exclusive processes to access the generalized parton distributions, semi-inclusive processes to study transverse momentum dependent distribution functions and inclusive spin structure functions, and resonance transition form factors at high $Q^2$ and with high precision. 
} 
%
\maketitle
\section{Introduction}
\label{intro}
The challenge of understanding nucleon electromagnetic structure still 
continues after more than five decades of experimental scrutiny. From the initial 
measurements of elastic form factors to the accurate determination of 
parton distributions through deep inelastic scattering (DIS), the
experiments have increased in statistical and systematic accuracy.  Only 
recently it was realized that the parton distribution functions
represent special cases of a more general, much more powerful, way to 
characterize the structure of the nucleon, the generalized parton 
distributions (GPDs)~\cite{Ji:1996nm,Ji:1996ek,Radyushkin:1996nd,Radyushkin:1997ki}.
  The GPDs are the Wigner quantum phase space 
distribution of quarks in the nucleon -- functions describing the 
simultaneous distribution of particles with respect to both position and 
momentum in a quantum-mechanical system, representing the closest analogue 
to a classical phase space density allowed by the uncertainty principle. 
In addition to the information about the spatial density (form factors) 
and momentum density (parton distribution), these functions reveal the 
correlation of the spatial and momentum distributions, {\it i.e.} how the 
spatial shape of the nucleon changes when probing quarks and gluons of 
different wavelengths.

The concept of GPDs has led to completely new methods of ``spatial imaging''
of the nucleon, either in the form of two-dimensional tomographic images 
(analogous to CT scans in medical imaging), or in the form of genuine 
three-dimensional images (Wigner distributions).  GPDs also allow us to 
quantify how the orbital motion of quarks in the nucleon contributes to the 
nucleon spin -- a question of crucial importance for our understanding of 
the ``mechanics'' underlying nucleon structure.  The spatial view of the 
nucleon enabled by the GPDs provides us with new ways to test dynamical 
models of nucleon structure. 

The mapping of the nucleon GPDs, and a detailed understanding of the
spatial quark and gluon structure of the nucleon, have been widely 
recognized as the key objectives of nuclear physics of the 
next decade. This requires a comprehensive program, combining results
of measurements of a variety of processes in electron--nucleon 
scattering with structural information obtained from theoretical studies, 
as well as with expected results from future lattice QCD simulations.

While GPDs, and also the recently introduced transverse momentum dependent 
distribution functions (TMDs), open up new avenues of research, the
traditional means of studying the nucleon structure through 
electromagnetic elastic and transition 
form factors, and through flavor- and spin-dependent parton distributions must also be 
employed with high precision to extract physics on the nucleon structure in the transition 
from the regime of quark confinement to the domain of asymptotic freedom. These 
avenues of research can be explored using the 12 GeV cw beam of the JLab 
upgrade with much higher precision than has been achieved ever before, 
and can help reveal some of the remaining secrets 
of QCD, such as the origin of confinement. Also, the high luminosity available 
will allow to explore the regime of extreme 
quark momentum, where a single quark carries 80\% or more of the proton's 
total momentum. To meet the requirements of high statistics measurements of relatively rare 
exclusive processes the equipment will be upgraded and include the {\tt CLAS12} 
large acceptance spectrometer. The main new features of {\tt CLAS12} over the current 
CLAS detector include a high operational luminosity of $10^{35}$cm$^{-2}$sec$^{-1}$, 
an order of magnitude increase over CLAS~\cite{clas}. Improved particle identification and event reconstruction will be achieved with 
additional threshold gas Cerenkov counter, improved timing resolution of the forward 
time-of-flight system, and a new central detector that uses a high-field solenoid magnet 
for particle tracking and the operation of a polarized targets. With these upgrades {\tt CLAS12}{ will be 
the workhorse for exclusive and semi-inclusive electroproduction experiments 
in the deep inelastic kinematics.

\begin{figure}[htb]
\resizebox{0.48\textwidth}{!}{%
  \includegraphics{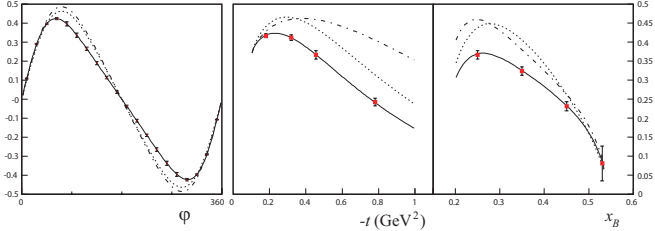}}
\caption{The beam spin asymmetry showing the DVCS-BH interference for 11 GeV beam 
energy~\cite{e12-06-119}. Left panel: $x=0.2$, 
$Q^2=3.3$GeV$^2$, $-t=0.45$GeV$^2$. Middle and right panels:  $\phi=90^{\circ}$, 
other parameters same as in left panel. Many other bins will be measured simultaneously. 
The curves represent various parameterizations within the VGG model~\cite{vgg}. Projected uncertainties are statistical.}
\label{fig:dvcs_alu_12gev}   
\end{figure}

\section{Generalized Parton Distributions and DVCS}
\label{sect:1}

It is well recognized~\cite{Ji:1996nm,Belitsky:2001ns,Burkardt:2002hr,Belitsky2004} that 
exclusive processes can be used to probe the GPDs and construct 2-dimensional 
 and 3-dimensional images of the quark content of the nucleon. Deeply virtual Compton scattering and
deeply virtual meson production are identified as the processes most suitable to 
map out at the twist-2 vector GPDs $H,~E$ and the axial GPDs ${\tilde H},~{\tilde E}$ in $x,~\xi,~t$, 
where $x$ is the momentum fraction of the 
struck quark, $\xi$ the longitudinal momentum transfer to the quark, and $t$ the 
transverse momentum transfer to the nucleon. Having access to a 3-dimensional image
of the nucleon (two dimensions in transverse space, one dimension in longitudinal 
momentum) opens up completely new insights into the complex structure of the
nucleon. In addition, GPDs carry information of more global nature. For example,
the nucleon matrix element of the energy-momentum tensor 
contains 3 form factors that encode information on the angular 
momentum distribution $J^q(t)$ of the quarks with flavor $q$ in transverse space, their 
mass-energy distribution $M_2^q(t)$, and their pressure and force 
distribution $d^q_1(t)$. How can we access these form factors? The only 
known process to directly measure them is elastic graviton scattering 
off the nucleon. However, today we know that these form factors also 
appear as moments of the vector GPDs~\cite{goeke2007}, thus offering
prospects of accessing these quantities through detailed mapping of GPDs.  The 
quark angular momentum in the nucleon is given by 
$$J^q(t) = 
\int_{-1}^{+1}dx x [H^q(x, \xi, t) + E^q(x, \xi, t)]~,$$ and the mass-energy and pressure distribution $$M_2^q(t) + 4/5d^q_1(t)\xi^2 
= \int_{-1}^{+1}dx x H^q(x, \xi, t)~.$$ The mass-energy and force-pressure distribution 
of the quarks are given by the second moment of GPD $\it{H}$, and their relative contribution is controlled by $\xi$. A separation of $M^q_2(t)$ and 
$d^q_1(t)$ requires measurement of these moments in a large range of 
$\xi$. The beam helicity-dependent cross section asymmetry is given 
in leading twist as 
$$ A_{LU} \approx \sin\phi[F_1(t)H + \xi(F_1+F_2)\tilde{H}]d\phi~, $$where
$\phi$ is the azimuthal angle between the electron scattering plane and the hadronic plane. The kinematically suppressed term with GPD $E$ is omitted. 
The asymmetry is mostly sensitive to the GPD $H(x=\xi,\xi,t)$. In a wide kinematics~\cite{clas-dvcs-1,clas-dvcs-3}  
the beam asymmetry $A_{LU}$ was measured at Jefferson Lab at modestly high $Q^2$, $\xi$, and $t$, and in a more limited kinematics~\cite{halla-dvcs} the cross section difference
$\Delta\sigma_{LU}$ was measured with high statistics. Moreover, 
a first measurement of the target asymmetry 
$A_{UL}=\Delta\sigma_{UL}/2\sigma$ was carried out~\cite{clas-dvcs-2}, where 
$$A_{UL} \approx \sin\phi[F_1\tilde{H} + \xi(F_1+F_2)H]~.$$  
The combination of $A_{LU}$ and $A_{UL}$ allows to separate GPD $H(x=\xi,\xi,t)$ and
$\tilde{H}(x=\xi,\xi,t)$.  
Using a transversely polarized target the asymmetry 
$$A_{UT} \approx \cos\phi\sin(\phi-\phi_s) [t/4M^2 (F_2H - F_1 E)] $$ can be measured, where $\phi_s$ is the azimuthal angle of the target polarization vector relative to the electron scattering plane. $A_{UT}$ 
depends in leading order on GPD $E$.

\begin{figure}
\resizebox{0.5\textwidth}{!}{%
  \includegraphics{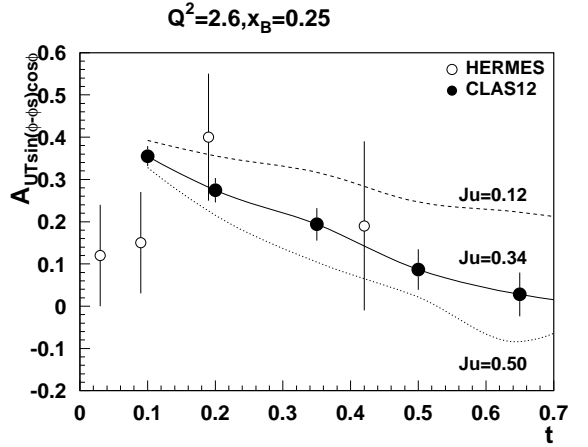}}
\caption{Projected transverse target asymmetry $A_{UT}$ for DVCS production off protons at 11 GeV beam energy.}
\label{fig:dvcs_aut_12gev}    
\end{figure}

The 12 GeV upgrade offers much improved possibilities to access GPDs. 
Figure~\ref{fig:dvcs_alu_12gev} shows the expected statistical precision of 
the beam DVCS asymmetry for some sample kinematics. Using a
polarized target one can also measure 
the  target spin asymmetries with high precision. Figure~\ref{fig:dvcs_aut_12gev} shows the expected statistical accuracy for one kinematics bin. A measurement of all 3 
asymmetries will allow a separate determination of GPDs 
$H,~\tilde{H}$ and $E$ at the above specified kinematics. Through a Fourier transformation
the t-dependence of GPD $H$ can be used to determine the $u-$quark distribution 
in transverse impact parameter space. Figure~\ref{fig:gpd_H} shows projected 
results.

Deeply virtual meson production will play an important role in disentangling 
the flavor- and spin-dependence of GPDs. For exclusive mesons only the longitudinal 
photon coupling in $\gamma^* p \rightarrow Nm$ allows direct access to GPDs through 
the handbag mechanism and must be isolated from the transverse coupling. Also, the 
dominance of the handbag mechanism must first be established at the upgrade energy.

\begin{figure}
\resizebox{0.5\textwidth}{!}{%
  \includegraphics{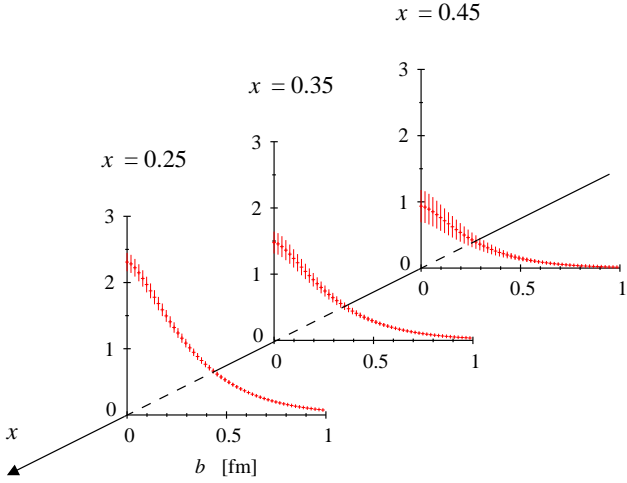}}
\caption{The u-quark distribution in transverse space as extracted from projected DVCS data with {\tt CLAS12}. }
\label{fig:gpd_H}    
\end{figure}

\section{Transverse momentum dependent parton distributions and SIDIS}
\label{sect:2}

 Semi-inclusive deep inelastic scattering
(SIDIS) studies, when a hadron is detected in coincidence with the scattered 
lepton that allows so-called ``flavor tagging'', provide more direct access to 
contributions from various quarks.  In addition, they give access to the 
transverse momentum distributions of quarks, not accessible in inclusive 
scattering.  Azimuthal distributions of final state particles in semi-inclusive 
deep inelastic scattering  provide access to the orbital motion of quarks and 
play an important role in the study of TMDs of 
quarks in the nucleon.
\begin{table}[h]
\begin{center}
\begin{tabular}{|c|c|c|c|} \hline\hline
N/q & U & L & T \\ \hline
 {U} & ${\bf f_1}$   & & ${ h_{1}^\perp}$ \\
\hline
 {L} & &${\bf g_1}$ &    ${ h_{1L}^\perp}$ \\
\hline
 {T} & ${ f_{1T}^\perp} $ &  ${ g_{1T}}$ &  ${ \bf h_1}$ \, ${ h_{1T}^\perp }$ \\
\hline\hline
\end{tabular}
\end{center}
\caption{\small{{Leading-twist transverse momentum-dependent distribution 
functions.  $U$, $L$, and $T$ stand for transitions of unpolarized, 
longitudinally polarized, and transversely polarized nucleons (rows) to 
corresponding quarks (columns).}}
\label{tab1}} 
\end{table}

TMD distributions (see Table~\ref{tab1}) describe transitions of a nucleon 
with one polarization in the initial state to a quark with another polarization 
in the final state. The diagonal elements of the table are the momentum, longitudinal and 
transverse spin distributions of partons, and represent well-known parton
distribution functions related to the square of the leading-twist, light-cone 
wave functions. Off-diagonal elements require non-zero orbital angular 
momentum and are related to the wave function overlap of $L$=0 and $L$=1 Fock 
states of the nucleon~\cite{Ji:2002xn}.  The chiral-even distributions 
$f_{1T}^\perp$ and $g_{1T}$ are the imaginary parts of the corresponding
interference terms, and the chiral-odd $h_1^\perp$ and $h_{1L}$ are the
real parts.  The TMDs $f_{1T}^\perp$ and  $h_{1}^\perp$, which are related to 
the imaginary part of the interference of wave functions for different orbital 
momentum states and are known as the Sivers and 
Boer-Mulders functions, and describe unpolarized quarks in the 
transversely polarized nucleon and transversely polarized quarks in the 
unpolarized nucleon respectively.  
The most simple mechanism that can lead to a Boer-Mulders function is a 
correlation between the spin of the 
quarks and their orbital angular momentum.  In combination with a final state 
interaction that is on average attractive, already a measurement of the sign 
of the Boer-Mulders function, would thus reveal the correlation between 
orbital angular momentum and spin of the quarks. 

Similar to GPDs, TMD studies will benefit from the higher energy and high 
luminosity at 12 GeV. A comprehensive program is in preparation with {\tt CLAS12} to study the new structure functions.
Examples of expected uncertainties~\cite{E12-07-107} for the Boer-Mulders asymmetry $A^{cos2\phi}_{UU}$ 
are presented in Fig.~\ref{fig:Boer-Mulders}. Projections of the 
Mulders function $h^u_{1L}$ for $u-$quarks from $\pi^+$ asymmetries $A_{UL}$ with {\tt CLAS12} are shown in Fig.~\ref{fig:h1lu11}, and compared with preliminary results from the {\tt CLAS} EG1 data set at 5.75 GeV beam energy.

\begin{figure}
\resizebox{0.5\textwidth}{!}{%
	\includegraphics{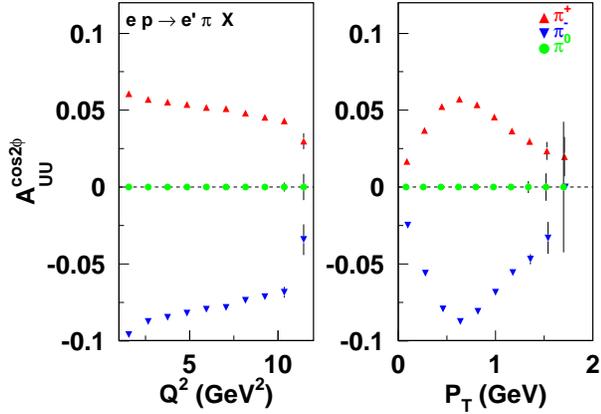}}
\caption{The $\cos2\phi$ moment (Boer-Mulders asymmetry) for pions
as a function of $Q^2$ and $P_T$ for $Q^2>2$~GeV$^2$ (right) with {\tt CLAS12} 
at 11~GeV from 2000~hours of running.  Values are calculated assuming
$H_1^{\perp u\rightarrow \pi^+}=-H_1^{\perp u\rightarrow \pi^-}$. Only statistical
uncertainties are shown.}
\label{fig:Boer-Mulders}    
\end{figure}

\begin{figure}[]
\resizebox{0.5\textwidth}{!}{%
	\includegraphics{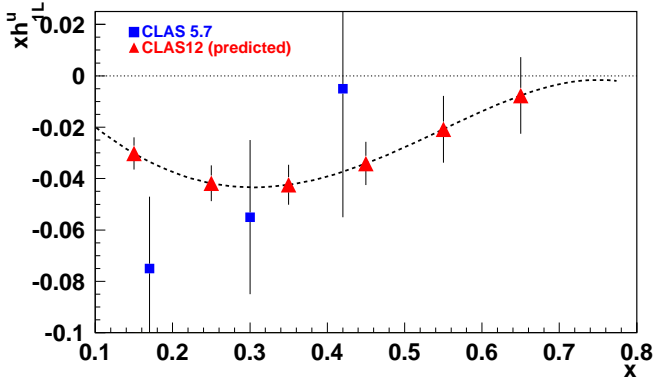}}
\caption{Projected data from {\tt CLAS12} for the chiral odd function $h^u_{1L}$ for $u$-quarks. }
\label{fig:h1lu11}    
\end{figure}

\section{Inclusive structure functions and moments}
\label{sect:3}
Polarized and unpolarized structure functions of the nucleon offer a
unique window on the internal quark structure of stable baryons.
The study of these structure functions provides insight into the two
defining features of QCD --- asymptotic freedom at small distances,
and confinement and non-perturbative effects at large distance scales.
After more than three decades of measurements at many accelerator
facilities worldwide, a truly impressive amount of data has been
collected, covering several orders of magnitude in both kinematic
variables $x$ and $Q^2$.
However, there are still important regions of the kinematic phase space
where data are scarce and have large errors and where significant
improvements are possible through experiments at Jefferson Lab with an
11~GeV electron beam.

\begin{figure}[bmt]
\resizebox{0.5\textwidth}{!}{%
  \includegraphics{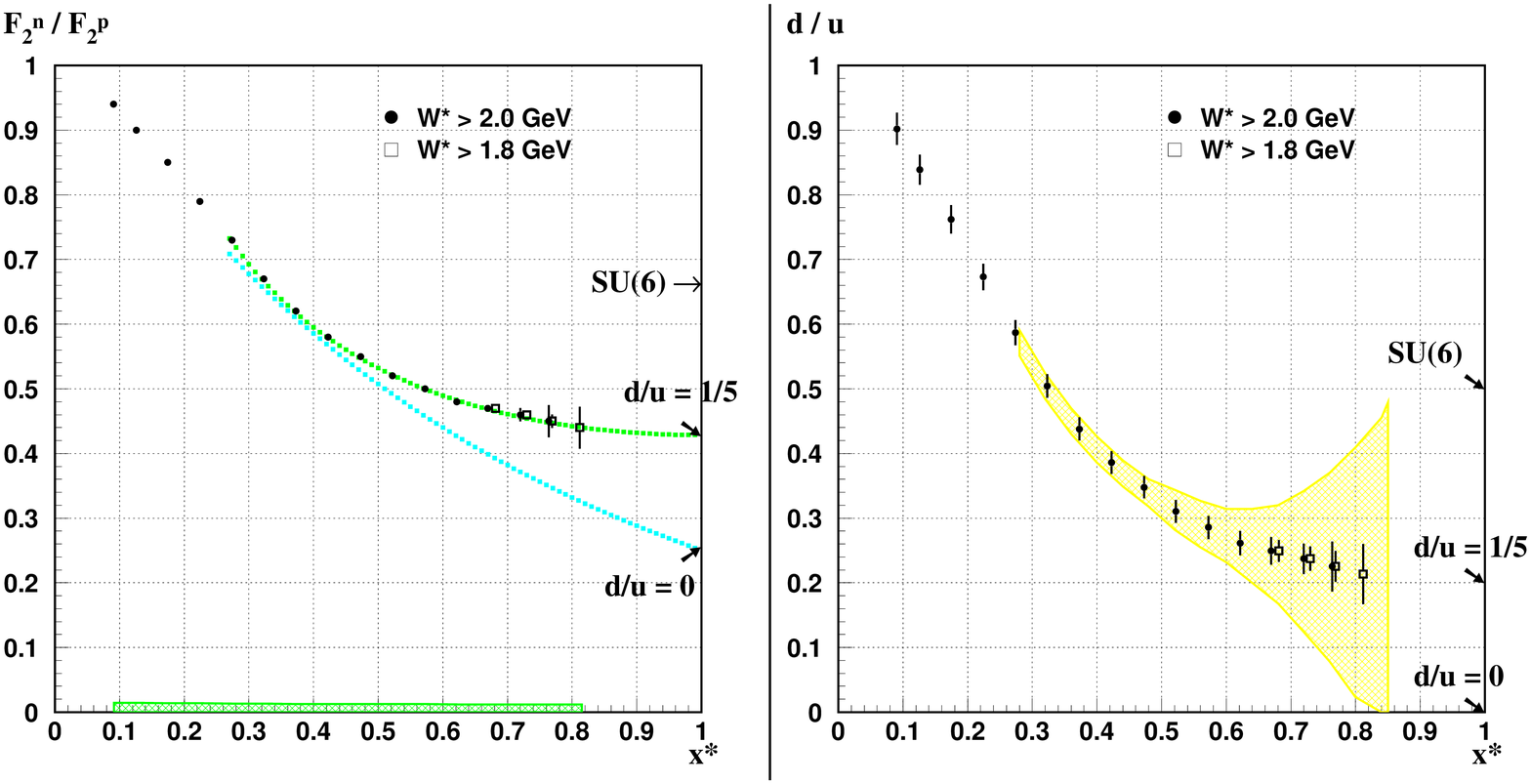}}
\caption{Projected data for the ratio $F^n_2/F^p_2$ (left) and $d/u$ (right) for 
11 GeV beam energy~\cite{bonus12_proposal}. The error bars in the right panel 
contain both statistical and 
systematic uncertainties. The yellow area shows the uncertainty of current data due to
poorly known nuclear corrections.}
\label{fig:F2nF2p}    
\end{figure}

One of the most interesting open questions is the behavior of the
structure functions in the extreme kinematic limit $x \rightarrow 1$.
In this region effects from the virtual sea of quark-antiquark pairs are 
suppressed, making this region simpler to model. This is also the region
where pQCD can make absolute predictions. However, the large $x$ 
domain is hard to reach because cross sections are kinematically suppressed, 
the parton distributions are small and final states interactions (partonic or 
hadronic) are large. First steps into the large $x$ domain became possible with 
5-6 GeV \cite{BONUS,zheng2004,vipuli2006,bosted2007}. The interest triggered by these
first results and the clear necessity to extend the program to larger $x$ 
provided one of the cornerstone of the JLab 12 GeV upgrade physics program.  

\begin{figure}
\resizebox{0.5\textwidth}{!}{%
  \includegraphics{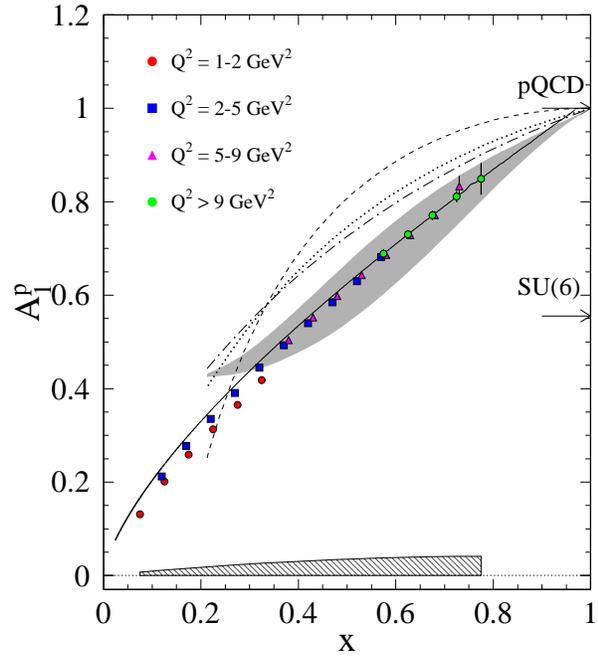}}
\caption{Anticipated results on $A_1^p$.The four different symbols represent four different $Q^2$ ranges.  The statistical uncertainty is given by the error bars while the systematic uncertainty is given by the shaded band. }
\label{fig:A1p}    
\end{figure}

\subsection{Valence quark structure and flavor dependence at large $x$.}
\label{sect:4}

The unpolarized structure function $F^p_2(x)$ has been mapped out in a 
large range of $x$ leading to precise knowledge of the quark distribution 
$u(x)$.  The corresponding structure function $F^n_2(x)$ is, however, well 
measured only for $x < 0.5$ as nuclear corrections, when using deuterium as 
a target, become large at large $x$ and are not well represented by 
Fermi motion. 
At JLab a new technique tested recently with {\tt CLAS} has been shown to be very
effective in reducing the nuclear corrections. The 
{\tt BONUS} experiment~\cite{BONUS} 
has recently taken data using a novel radial TPC with GEM readout as detector for 
the low-energy spectator proton in the reaction $en(p_s)\rightarrow ep_sX$.
Measurement of the spectator proton for momenta as low as 70 MeV/c and at large 
angles allows
to minimize poorly known nuclear corrections at large $x$. The 
techniques can be used with {\tt CLAS12} at 12~GeV to accurately determine 
the ratio $d(x)/u(x)$ to much larger $x$ values. 
Figure~\ref{fig:F2nF2p} shows the projected data for $F^n_2(x)/F^p_2(x)$ and $d(x)/u(x)$. 
A dramatic improvement can be achieved at large $x$.

\subsection{Spin structure functions and parton distributions}
\label{sect:5}

JLab PAC30 also approved E12-06-109 \cite{EG12} which will, in particular, study polarized parton distributions at large $x$. Using standard detection equipment, a redesigned polarized target adapted to {\tt CLAS12} and 30 (50)~days of running on a longitudinally polarized NH$_3$ (ND$_3$) target, high precision measurements can be achieved as shown in Fig.~\ref{fig:A1p}. These data will disentangle models in the large-$x$ region.  While the results shown in Fig.~\ref{fig:A1p} are with a $W>2$~GeV constraint, hadron-parton duality studies will tell us by how much this constraint can be relaxed, possibly increasing the $x$ range up to 0.9.  The expected accuracy for $(\Delta d+ \Delta\bar{d})/(d+\bar{d})$ is shown in Fig.~\ref{fig:ddodxpctd}.
  
\begin{figure}
\resizebox{0.5\textwidth}{!}{%
  \includegraphics{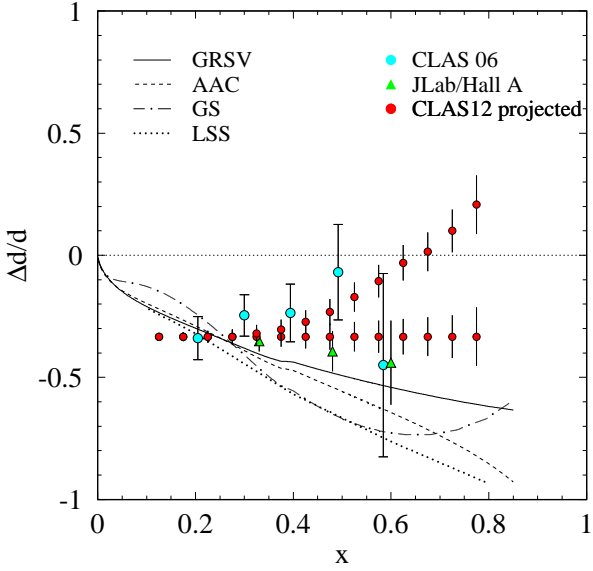}}
\caption{Expected results for $(\Delta d+ \Delta\bar{d})/(d+\bar{d})$. The central values of the data are following two arbitrary curves to demonstrate how the two categories of predictions, namely the ones that predict $\Delta d/d$ stays negative (LO and NLO analyses of polarized DIS data: GRSV, LSS, AAC, GS, statistical model, and a quark-hadron duality scenario) and the ones predicting $\Delta d/d \to 1$ when $x \to 1$ (leading order pQCD and a quark-hadron duality scenario).}
\label{fig:ddodxpctd}    
\end{figure}

\subsection{Global Fit of Polarized Parton Distributions}
\label{sect:6}
The large window opened by the 12 GeV upgrade over the DIS domain will permit 
constraints of global fits of the parton distributions.  JLab data at lower 
energies had already unique impact
at large $x$. The improvement from the 12-GeV upgrade is 
also significant at low and moderate $x$, noticeably for the polarized gluon 
distribution $\Delta G$.  To demonstrate the precision 
achievable with the expected {\tt CLAS12} data, we have plotted in 
Fig.~\ref{fig:pPDFs_exp} an analysis of the impact on NLO analyses of the 
polarized gluon distribution~\cite{LSS2007}.  A dramatic 
improvement can be achieved with the expected data from the {\tt CLAS12} 
proposal E12-06-109~\cite{EG12}.  We emphasize that the data will not only 
reduce the error band on $\Delta G$, but will likely allow a more detailed 
modeling of its $x$-dependence. Significant improvements are expected for the quark 
distributions as well.

\subsection{Moments of spin structure functions.}
\label{sect:7}
Moments of structure functions provide powerful insight into nucleon 
structure.  Inclusive data at JLab have permitted evaluation of the moments 
at low and intermediate $Q^2$~\cite{fatemi2003,yun2003,chen2004}.  
With a maximum beam energy of 6~GeV, however, the measured strength of the 
moments becomes rather limited for $Q^2$ greater than a few GeV$^2$. The 
12-GeV upgrade removes this problem and allows for measurements to higher 
$Q^2$. 

\begin{figure}
\resizebox{0.5\textwidth}{!}{%
  \includegraphics{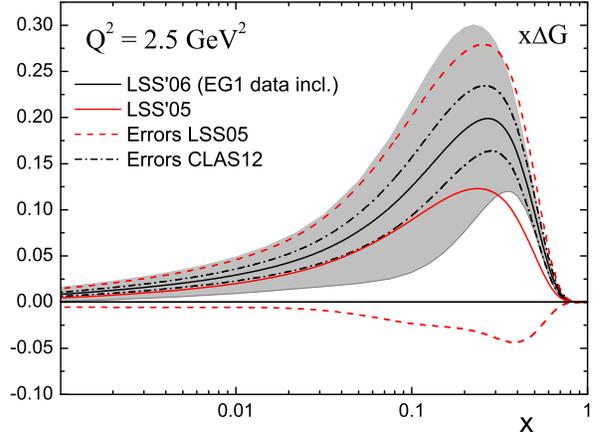}}
\caption{Expected uncertainties for $x\Delta{G}$. The black solid curve shows the central value of the present analysis that includes CLAS EG1 data. The dashed-dotted lines give the error band when the expected {\tt CLAS12} data are included in the LSS QCD analysis.}
\label{fig:pPDFs_exp}    
\end{figure}

Moments of structure functions can be related to the nucleon static properties 
by sum rules. At large $Q^2$ the Bjorken sum rule relates $\int g_1^{p-n} dx$ 
to the nucleon axial charge~\cite{Bjorken:1966jh}. 
Figure~\ref{fig:expect} shows the expected precision on $\Gamma_1^p$. Published 
results and preliminary results from EG1b are also displayed for comparison. 
The hatched blue band corresponds to the systematic uncertainty 
on the EG1b data points.  The red band indicates the estimated systematic 
uncertainty from {\tt CLAS12}.  The systematic uncertainties for EG1 and 
{\tt CLAS12} include the estimated uncertainty on the unmeasured DIS part 
estimated using the model from Bianchi and Thomas~\cite{Thomas:2000pf}.  As 
can be seen, moments can be measured up to $Q^2$=6~GeV$^2$ with a statistical 
accuracy improved several fold over that of the existing world data.

\begin{figure}
\resizebox{0.5\textwidth}{!}{%
  \includegraphics{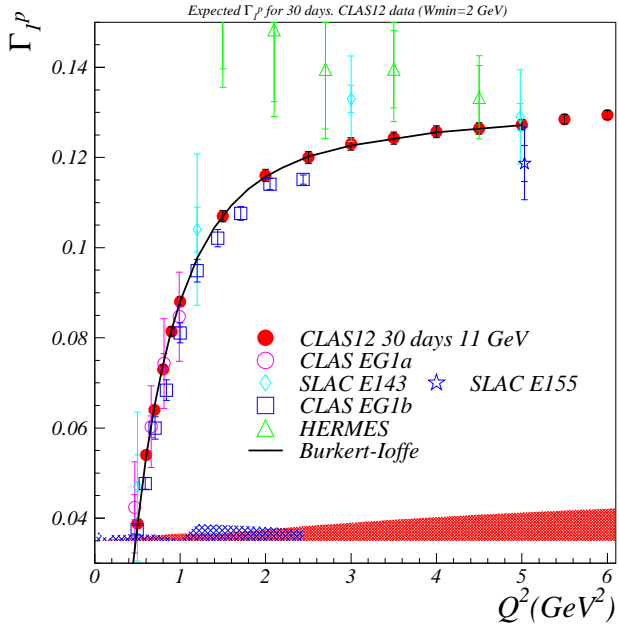}}
\caption{Left plot: expected precision on $\Gamma_1^p$ for {\tt CLAS12}
and 30~days of running.  {\tt CLAS} EG1a~\cite{fatemi2003,yun2003} data 
and preliminary results from EG1b are shown for comparison.  The data and 
systematic uncertainties include estimates of the unmeasured DIS 
contribution.  HERMES~\cite{Airapetian:2002wd} data, and E143~\cite{Abe:1998wq} 
and E155 data~\cite{Anthony:2000fn} from SLAC are also shown (including DIS 
contribution estimates).  The model is from Burkert and Ioffe
\cite{Burkert:1992tg,Burkert:1993ya}.}
\label{fig:expect}    
\end{figure}

Finally, moments in the low ($\simeq$ 0.5 GeV$^2$) to moderate 
($\simeq$4~GeV$^2$) $Q^2$ range enable us to extract higher-twist parameters,
which represent correlations between quarks in the nucleon.  This extraction 
can be done by studying the $Q^2$ evolution of first moments~\cite{Osipenko2005,Chen:2005td}.
Higher twists have been consistently found to have, overall, a surprisingly 
smaller effect than expected.  Going to lower $Q^2$ enhances the higher-twist 
effects but makes it harder to disentangle a high twist from the yet higher 
ones.  Furthermore, the uncertainty on $\alpha _s$ becomes prohibitive at low 
$Q^2$.  Hence, higher twists turn out to be hard to measure, even at the 
present JLab energies.  Adding higher $Q^2$ to the present JLab data set 
removes the issues of disentangling higher twists from each other and of the 
$\alpha _s$ uncertainty.  The smallness of higher twists, however, requires 
statistically precise measurements with small point-to-point correlated 
systematic uncertainties.  Such precision at moderate $Q^2$ has not been 
achieved by the experiments done at high energy accelerators, while JLab at 
12~GeV presents the opportunity to reach it considering the expected 
statistical and systematic uncertainties of E12-06-109.  The total 
point-to-point uncorrelated uncertainty on the twist-4 term for the Bjorken 
sum, $f_2^{p-n}$, decreases by a factor of 5.6 compared to results obtained in
Ref.~\cite{Deur:2004ti}. 

\begin{figure}[]
\resizebox{0.45\textwidth}{!}{%
  \includegraphics{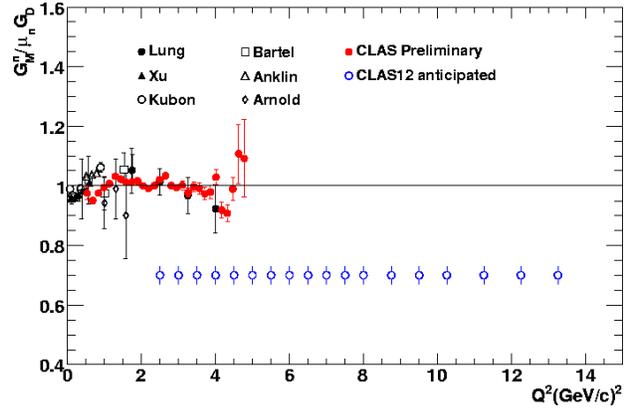}}
\caption{The magnetic form factor for the neutron. The existing data, and projected uncertainties at 12 GeV with {\tt CLAS12} (blue open circles).}
\label{fig:gmn}       
\end{figure}

\section{Nucleon form factors and resonance transitions at short distances}
\label{sect:8}
  
The most basic observables that reflect the composite nature of the 
nucleon are its electromagnetive form factors. Historically the first 
direct indication that the nucleon is not elementary came from 
measurements of these quantities in elastic $ep$ scattering~\cite{HOF}. 
The electric and magnetic form factors characterize the distributions of 
charge and magnetization in the nucleon as a function of spatial resolving 
power. The transition form factors reveal the nature of the excited states of the 
nucleon.  Further, these quantities can be described and related to other 
observables through the GPDs.

Measurements of the elastic form factors will remain an important aspect 
of the physics program at 12 GeV, and will be part of the program in other 
experimental Halls at JLab. The magnetic form factor of the 
neutron, as well as the transition form factors  for several prominent 
resonances require special experimental setups for which {\tt CLAS12} 
is suited best. Figure ~\ref{fig:gmn} shows the current data as well as the extension 
in $Q^2$ projected for the 12 GeV program with {\tt CLAS12}. 

\begin{figure}
\resizebox{0.4\textwidth}{!}{%
  \includegraphics{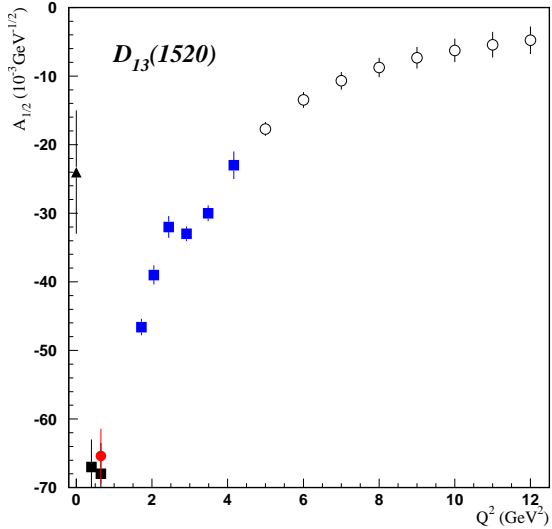}}
\caption{The transverse photocoupling amplitude $A_{1/2}$ for the 
$N(1520)D_{13}$ resonance.
The blue full squares are preliminary data from CLAS. The open 
circles represent projected results with {\tt CLAS12} at 12 GeV.}
\label{fig:d13_high}       
\end{figure}

Nucleon ground and excited states represent different eigenstates of the Hamiltonian,
therefore to understand the interactions underlying nucleon formation from fundamental 
constituents, the structure of both the ground state and the excited states must be studied. The current $N^*$ program at Jlab has already generated results for the transition 
form factors at $Q^2$ up to 6 GeV$^2$ for the $\Delta(1232)$~\cite{frolov1999,joo2002,ungaro2006}, and up to 4~GeV$^2$ for the $N(1535)S_{11}$~\cite{armstrong1999,thompson2001,denizli2007}. The most recent results~\cite{park2007,aznauryan2007}  
on the transition form factors for the Roper resonance $N(1440)P_{11}$ for $Q^2$ up to 4.5~GeV$^2$, have 
demonstrated the sensitivity to the degrees 
of freedom that are effective in the excitation of particular states. 
The JLab energy upgrade will allow us to probe 
resonance excitations at much higher $Q^2$, where the relevance of elementary 
quarks in the resonance formation may become evident through the approach
to asymptotic scaling. Figure~\ref{fig:d13_high} shows projected $Q^2$ dependence 
of the $A_{1/2}$ transition amplitude for the $N(1520)D_{13}$ resonance obtained from single pion production. Higher mass resonances may be efficiently measured in double-pion processes~\cite{ripani2003,mokeev} such as $ep\rightarrow ep\pi^+\pi^-$.    

\section{Conclusions}
\label{summary}

The JLab energy upgrade and the planned new experimental equipment are well 
matched to an exciting scientific program aimed at studies 
of the complex nucleon structure in terms of the newly discovered 
longitudinal and transverse momentum dependent parton distribution 
functions, the GPDs and TMDs. 
They provide fundamentally new insights in the complex multi-dimensional 
structure of the nucleon. In addition, the high precision afforded by the 
high luminosity and the large acceptance detectors, and the development of 
novel techniques to measure scattering off nearly free neutrons, will allow 
the exploration of phase space domains with extreme conditions that could not be 
studied before.

\vspace{0.3cm}

\noindent{\bf Acknowledgment}

I am grateful to members of the CLAS collaboration who contributed to the development of the exciting physics program for the JLab upgrade to 12 GeV, and the {\tt CLAS12} detector. Much of the material in this report is taken from the {\tt CLAS12} Technical Design Report Version 3, October 2007~\cite{clas12_tdr}.  

This work was supported in part by the U.S. Department of Energy and the National Science Foundation, the French Commisariat {\'{a}} l'Energie Atomique, the Italian Instituto Nazionale di Fisica Nucleare, the Korea Research Foundation,  and a research grant of the Russian Federation. The Jefferson Science Associates, LLC, operates Jefferson Lab under contract DE-AC05-060R23177.


\begin{thebibliography}{}
\bibitem{clas12_tdr} The complete document may be obtained from the author.
\end{thebibliography}
\end{document}